\documentclass{PoS}
\usepackage{refmerge}
\RequirePackage{xspace}
\usepackage{relsize}
\def\babar{\mbox{\slshape B\kern-0.1em{\smaller A}\kern-0.1em
    B\kern-0.1em{\smaller A\kern-0.2em R}}}
\mathchardef\Upsilon="7107
\def\Y#1S{\ensuremath{\Upsilon{(#1S)}}\xspace}
\def\FourS {\Y4S}
\def\epem{\ensuremath{e^+e^-}\xspace}

\def\Vus  {\ensuremath{|V_{us}|}\xspace}

\def\mtau{\ensuremath{\tau}\xspace}
\def\mmu        {\ensuremath{\mu}\xspace}
\def\ckm{CKM}
\def\ms{\ensuremath{m_s}\xspace}
\def\Klthree{\ensuremath{K{\ell}3}\xspace}
\def\Kltwo{\ensuremath{K{\ell}2}\xspace}
\def\tautau{\ensuremath{\tau^+\tau^-}\xspace}
\newcommand{\eett}   {\ensuremath{e^+e^- \to \tautau}\xspace}
\def\nb         {\ensuremath{{\rm \,nb}}\xspace}

\newcommand{\roots}        {\ensuremath{\sqrt{s}}\xspace}
\newcommand{\gev}{\ensuremath{\mathrm{\,Ge\kern -0.1em V}}\xspace}
\newcommand{\mev}{\ensuremath{\mathrm{\,Me\kern -0.1em V}}\xspace}
\newcommand{\mevcc}{\ensuremath{{\mathrm{\,Me\kern -0.1em V\!/}c^2}}\xspace}
\def\kk       {\mbox{\tt KK2f}\xspace}

\newcommand{\tautophip} {\ensuremath{\tau^- \to \phip \nu_\tau}\xspace}

\newcommand{\tautophik} {\ensuremath{\tau^- \to \phik   \nu_\tau}\xspace}

\newcommand{\kkk} {\ensuremath{K^-K^-K^+}\xspace}
\newcommand{\phip} {\ensuremath{\phi\pi^-}\xspace}
\newcommand{\phik} {\ensuremath{\phi K^-}\xspace}
\def\piz   {\ensuremath{\pi^0}\xspace}

\def\Kbar  {\kern 0.2em\overline{\kern -0.2em K}{}\xspace}
\def\Kb    {\ensuremath{\Kbar}\xspace}
\def\Kz    {\ensuremath{K^0}\xspace}
\def\Kzb   {\ensuremath{\Kbar^0}\xspace}
\def\KzKzb {\ensuremath{\Kz \kern -0.16em \Kzb}\xspace}
\def\Kp    {\ensuremath{K^+}\xspace}
\def\Km    {\ensuremath{K^-}\xspace}

\def\KpKm  {\ensuremath{\Kp \kern -0.16em \Km}\xspace}
\def\KS    {\ensuremath{K^0_{\scriptscriptstyle S}}\xspace}

\def\Kstarm  {\ensuremath{K^{*-}}\xspace}

\def\BR         {{\ensuremath{\cal B}\xspace}}
\newcommand{\BRtauphip}    {\ensuremath{\BR(\tautophip)}\xspace}
\newcommand{\BRtauphik}    {\ensuremath{\BR(\tautophik)}\xspace}

\title{Measurement of {\boldmath $\Vus$} using hadronic tau decays from \babar\ \& Belle}

\ShortTitle{\Vus from \mtau decays}

\author{\speaker{Swagato Banerjee}%
        \thanks{For \babar\ \& Belle collaborations.}\\
        University of Victoria, P.O. Box 3055, Victoria, B.C., CANADA  V8W 3P6.\\
        E-mail: \email{swaban@slac.stanford.edu}}

\abstract{ We report on measurements of branching fractions for several hadronic tau
           decays to final states with kaons, which can be used to determine
           the strange quark mass and the element \Vus of the
           Cabibbo-Kobayashi-Maskawa quark-mixing matrix.  
           The results are obtained from data collected with the \babar\ and Belle 
           detectors at the PEP-II and KEKB asymmetric-energy \epem colliders
           at SLAC and KEK, respectively, both operating at center-of-mass energies
           near the \FourS resonance.}

\FullConference{KAON International Conference\\
                 May 21-25 2007\\
                 Laboratori Nazionali di Frascati dell'INFN, Rome, Italy.

}

\begin{document}

\section{Introduction}

The weak interaction universality between quarks was asserted
by Cabibbo with the introduction of a mixing angle 
between the first and second quark generations~\cite{Cabibbo:1963yz}.
Three-generation quark mixing between the mass eigenstates and
the flavor eigenstates is described by the unitary Cabibbo-Kobayashi-Maskawa (\ckm)
matrix~\cite{Kobayashi:1973fv}.  

The magnitude of the largest off-diagonal element of this \ckm\ matrix, \Vus,
has historically been measured from three-body kaon
(\Klthree) and hyperon decays~\cite{Leutwyler:1984je,Cabibbo:2003ea}.
\Vus has also been extracted from a comparison of the radiative inclusive rates
of two-body kaon (\Kltwo) and pion decays, along with lattice-QCD
results of the meson decay constants~\cite{Marciano:2004uf}. 
Recently, the analysis of flavor-breaking sum rules has shown that 
inclusive measurements of the strange spectral function, 
obtained from hadronic \mtau decays having net strangeness
of unity in the final state, can provide a direct determination of 
\Vus and the strange quark mass, \ms~\cite{Gamiz:2004ar,Chen:2001qf,Kambor:2000dj,Pich:1999hc,Narison:1999mv,Maltman:2006ic,Maltman:2006mr}.

Present generation $B$ factories, \babar\ and Belle, also serve as \mtau factories
thanks to the large \mtau pair production cross-section 
$\sigma_{\eett} = 0.919 \pm 0.003 \nb$~\cite{Banerjee:2007is},
as determined using the \kk Monte Carlo (MC) generator~\cite{Ward:2002qq}
at a center-of-mass (CM) energy of \roots =  10.58 \gev.
Using the world's largest sample of hadronic \mtau decays collected with the \babar\ and Belle detectors
at the PEP-II and KEKB asymmetric-energy \epem colliders, 
measurement of \ms and \Vus can be performed with unprecedented precision~\cite{Gamiz:2006xx,Maltman:2007pr}. 

Study of the strange spectral function from these \mtau data is still in progress.
However, using the updated knowledge of $\ms(2\gev) = 94 \pm 6 \mevcc$ from lattice calculations~\cite{Jamin:2006tj}, 
\Vus can be extracted with relatively small theoretical uncertainties~\cite{Davier:2005xq,Gamiz:2006xx,Gamiz:2007qs},
using available measurements of branching fractions of all \mtau decays into final states containing an odd number of kaons.

\section{Hadronic \mtau decays}

Here we report on the recent \babar~\cite{Aubert:kpi0,Aubert:hhh} and Belle~\cite{Epifanov:2007rf} measurements:
\begin{eqnarray*}
\BR(\tau^- \to K^- \piz \nu_\tau) & =  & (0.416 \pm 0.003 \pm 0.018)\%~\cite{Aubert:kpi0},\\
\BR(\tau^- \to \pi^- \pi^- \pi^+ \nu_\tau) & = & (8.83 \pm 0.01 \pm 0.13)\%~\cite{Aubert:hhh},\\
\BR(\tau^- \to  K^-  \pi^- \pi^+ \nu_\tau) & = & (0.273\pm 0.002\pm 0.009)\%~\cite{Aubert:hhh},\\
\BR(\tau^- \to  K^-  \pi^-  K^+ \nu_\tau)  & = & (0.1346\pm 0.0010 \pm 0.0036)\%~\cite{Aubert:hhh},\\
\BR(\tau^- \to  K^-   K^-   K^+ \nu_\tau)  & = & (1.58\pm 0.13 \pm 0.12)\times 10^{-5}~\cite{Aubert:hhh}~\mathrm{and}\\
\BR(\tau^- \to \KS  \pi^- \nu_\tau) & = & (0.404 \pm 0.002 \pm 0.013)\%~\cite{Epifanov:2007rf},
\end{eqnarray*}
where the uncertainties are statistical and systematic, respectively,
and the charge-conjugate modes are implied.
These results are more precise than the previously published measurements~\cite{Yao:2006px}.

The vector $K^{\star}(892)^-$ resonance is expected to nearly saturate the $(K\pi)^-$ final state.
The Belle analysis of the $\KS\pi^-$ invariant mass spectrum reveals contributions
from the vector $K^{\star}(892)^-$ resonance as well as other states~\cite{Epifanov:2007rf}. 
For the first time, the $K^{\star}(892)^-$ mass and width have been measured in \mtau decays: 
$m(K^{\star}(892)^−) = (895.47 \pm 0.20 (stat) \pm 0.44(syst) \pm 0.59(mod)) \mevcc$,
$\Gamma(K^{\star}(892)^−) = (46.2 \pm 0.6(stat) \pm 1.0(syst) \pm 0.7(mod)) \mev$. 
The $K^{\star}(892)^-$ mass is significantly different from the current world-average value
of $891.66 \pm 0.26\mevcc$ ~\cite{Yao:2006px}.

We also report on the first measurement of
$\BRtauphik =  (4.05\pm 0.25 \pm 0.26)\times 10^{-5}$ by the Belle experiment~\cite{Inami:2006vd}, 
which is consistent with the new \babar\ measurement of
$\BRtauphik =  (3.39\pm 0.20 \pm 0.28)\times 10^{-5}$,
and show that the $\tautophik$ decay saturates the \kkk final state~\cite{Aubert:hhh}.
The first measurement of 
$\BRtauphip = (3.42\pm 0.55 \pm 0.25)\times 10^{-5}$ 
has been performed by the \babar\ experiment~\cite{Aubert:hhh}, 
which provides an interesting laboratory to study OZI suppression~\cite{LopezCastro:1996xh},
because the \tautophik decay, with a comparable rate, is not OZI suppressed.

\section{\boldmath $\Vus$ from $\mtau$ decays} 

The hadronic width of the \mtau is normalized as: 
$R_\tau = \Gamma[\tau^- \to \mathrm{hadrons}^- \nu_\tau (\gamma)]/
         \Gamma[\tau^- \to e^- \bar{\nu}_e \nu_\tau (\gamma)]$.
The spectral moments are defined as:
$R^{kl}_\tau = \int^1_0 dz (1-z)^k z^l \frac{dR_\tau}{dz}$, 
where $z = \frac{q^2}{m^2_\tau}$ is the square of the scaled invariant mass of the hadronic system.
The flavor-breaking term 
$\delta R^{kl}_\tau = \frac{R^{kl}_{\tau,\rm{non-strange}}}{|V_{ud}|^2} - 
                     \frac{R^{kl}_{\tau,\rm{strange}}}{|V_{us}|^2}$
is sensitive to \ms. 
This term for the $kl = 00$ moment has the smallest theoretical uncertainty (given in parentheses):
$\delta R^{00}_{\tau,th} = 0.1544 (37) + 9.3 (3.4) \ms^2 + 0.0034 (28) = 0.240 (32)$~\cite{Gamiz:2006xx,Jamin:2006tj}.

We can then measure
$|V_{us}| = \sqrt{R^{00}_{\tau,\rm{strange}}/\left[\frac{R^{00}_{\tau,\rm{non-strange}}}{|V_{ud}|^2} -  \delta R^{00}_{\tau,th}\right]}$
from the measured \mtau branching fraction into strange final states,
where we use $|V_{ud}| = 0.97377 \pm 0.00027$~\cite{Yao:2006px}.
The modest 13\% error on $\delta R^{00}_{\tau,th}$ gives a relatively small contribution to the error on \Vus\ ~\cite{Gamiz:2007qs}.

Here, we use the values of \mtau branching fractions as estimated in Reference~\cite{Davier:2005xq}.
The electronic branching fraction $\BR^{\rm{uni}}_e  = (17.818 \pm 0.032)\%$
is obtained by averaging the direct measurements of 
the electronic and muonic branching fractions and the lifetime of the \mtau lepton.
Using lepton universality, the total hadronic \mtau branching fraction is
$\BR_{\rm{had}} = 1 - 1.97257 \BR^{\rm{uni}}_e  = (64.853 \pm 0.063)\%$,
and the total \mtau hadronic width is $R_\tau = 3.640 \pm 0.010$.
The non-strange width is $R_{\rm{non-strange}} = R_{\tau} - R_{\rm{strange}}$,
where the branching fractions into strange final states 
are listed in Table~\ref{table:results}. 

\begin{table}[!htbp]
\caption[.]{\mtau branching fractions into strange final states from~\cite{Davier:2005xq},
            averaged with results from~\cite{Aubert:kpi0,Aubert:hhh,Epifanov:2007rf}
            along a scale factor $S$ following the PDG prescription~\cite{Yao:2006px},
            sum up to $(28.44\pm 0.74)\times10^{-3}$ as shown in this table.
            If the ${\cal{B}}(\tau^-\to K^-\nu)$ is replaced with $(7.15\pm0.03)\times10^{-3}$,
            the sum becomes $(28.78\pm0.71)\times10^{-3}$.
            }
\label{table:results}
\begin{center}
\setlength{\tabcolsep}{0.1pc}
\begin{tabular*}{.99\textwidth}{@{\extracolsep{\fill}}l@{\hspace*{2mm}}c@{\hspace*{2mm}}c} 
\hline\noalign{\smallskip}
 Mode & \BR $(10^{-3})$~\cite{Davier:2005xq}& Updated \BR $(10^{-3})$ with results from~\cite{Aubert:kpi0,Aubert:hhh,Epifanov:2007rf}\\
\noalign{\smallskip}\hline\noalign{\smallskip}
  $K^-$                         & $6.81 \pm 0.23$ & \\
  $K^- \pi^0$                   & $4.54 \pm 0.30$ & Average with $4.16 \pm 0.18$ $\Rightarrow$ $4.26 \pm 0.16$ $(S=1.0)$\\
  $\Kzb \pi^-$                  & $8.78 \pm 0.38$ & Average with $8.08 \pm 0.26$ $\Rightarrow$ $8.31 \pm 0.28$ $(S=1.3)$\\
  $K^- \pi^0 \pi^0$             & $0.58 \pm 0.24$ & \\
  $\Kzb \pi^- \pi^0$            & $3.60 \pm 0.40$ & \\
  $K^- \pi^+ \pi^-$             & $3.30 \pm 0.28$ & Average with $2.73 \pm 0.09$ $\Rightarrow$ $2.80 \pm 0.16$ $(S=1.9)$\\
  $K^- \eta$                    & $0.27 \pm 0.06$ & \\
  $(\Kb 3\pi)^-$ (estimated)    & $0.74 \pm 0.30$ & \\
  $K_1(1270)^- \to K^- \omega$  & $0.67 \pm 0.21$ &\\
  $(\Kb 4\pi)^-$ (estimated) and $\Kstarm \eta$     
                                & $0.40 \pm 0.12$ & \\
\noalign{\smallskip}\hline\noalign{\smallskip}
 Sum                           &$29.69 \pm 0.86$ & Updated Estimate: $28.44 \pm 0.74$ \\
\noalign{\smallskip}\hline
\end{tabular*}
\end{center}
\end{table}

The results for the $K^- \pi^0$, $K^- \pi^+ \pi^-$ and $\Kzb \pi^-$ branching fractions in~\cite{Davier:2005xq} 
have been averaged with the results presented here~\cite{Aubert:kpi0,Aubert:hhh,Epifanov:2007rf},
where the errors include a scale factor $S$ following the PDG prescription~\cite{Yao:2006px}.
The total branching fraction into strange final states has also been calculated by replacing 
$\BR(\tau^- \to K^- \nu_\tau)$ with $(7.15 \pm 0.03) \times 10^{-3}$,
obtained from theoretical predictions using the much better known
$K^- \to \mu^- \nu_\mu (\gamma)$ decay rate and assuming \mtau-\mmu universality.

The updated value is $\Vus =  0.2157 \pm 0.0031$ using measured \mtau branching fractions alone,
and $\Vus = 0.2171 \pm 0.0030$ using the predicted $\tau^- \to K^- \nu_\tau$ branching fraction as well.
The uncertainties are dominated by the $\sim$ 2\% experimental errors on the measurements
of the \mtau branching fractions.
Some of the uncertainties have been reduced almost by a factor of two,
because of the new \babar\ and Belle results.
In the near future, we expect to reduce the uncertainty on the strange branching fractions by an additional factor of two,
corresponding to a precision of 0.7\% on \Vus.

\section{Summary}

\begin{figure}[!hbtp]
\begin{center}
\caption{Comparison of different estimates of \Vus.}
\resizebox{.8\columnwidth}{.44\textheight}{%
\includegraphics{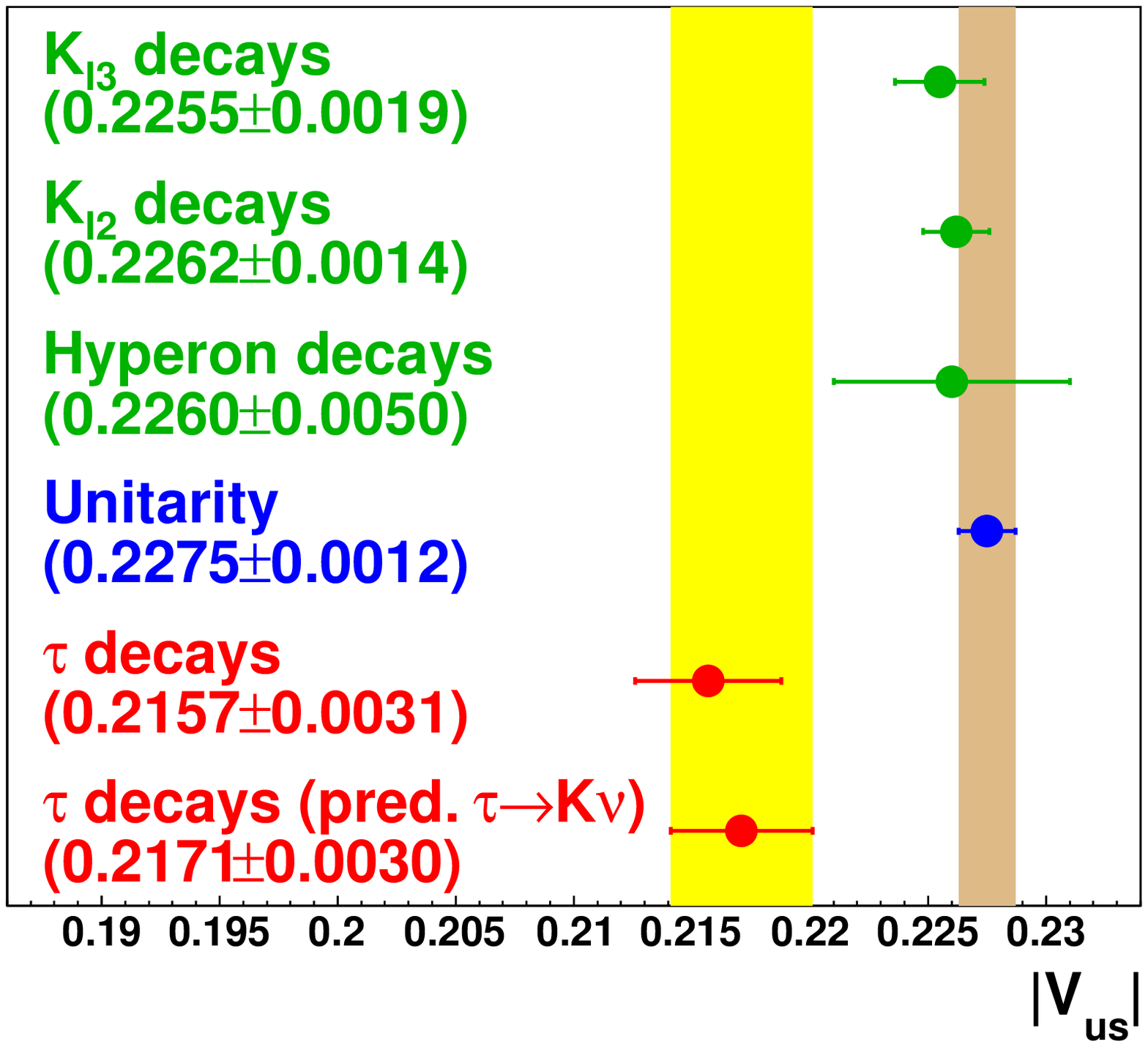}}
\label{fig:compare}
\end{center}
\end{figure}

In Figure~\ref{fig:compare}, updated estimates of \Vus from \mtau decays are compared
 with those extracted from \Klthree~\cite{Palutan:Kaon07}, 
\Kltwo~\cite{Follana:2007uv} and hyperon decays~\cite{Jamin:MoriondEW07}.
The values obtained from \mtau decays with and without the predicted $\tau^- \to K^- \nu_\tau$ branching fraction
are lower than the estimate of $\Vus =  0.2275 \pm 0.0012$
obtained by using the unitarity constraint from the value of $|V_{ud}|$, mentioned above,
by $3.2 \sigma$ and $3.5 \sigma$, respectively.
Possible implications for new physics due to this departure 
from the unitarity constraint are discussed in Reference~\cite{Marciano:Kaon07}.

\end{document}